\documentclass[
 reprint,
 amsmath,amssymb,
 aps,
 showkeys
]{revtex4-1}

\usepackage{graphicx}% Include figure files
\usepackage{dcolumn}% Align table columns on decimal point
\usepackage{bm}% bold math
\usepackage{hyperref}

\begin{document}

\title{Analytic theory of coupled waveguide transformation under irregular  perturbations}

\author{Mikhail Yu.~Saygin}
 \email{saygin@physics.msu.ru}
 
\affiliation{
 Quantum Technologies Center and Faculty of Physics, M.V.Lomonosov Moscow State University, GSP-1, Leninskie gory, Moscow 119991  Russian Federation
}

%\date{\today}

\begin{abstract}
Coupled waveguides are convenient at implementing useful optical transformations.
In this letter, we investigate the uncertainty of transformation by two coupled waveguides 
that stems from  the perturbations of waveguide's mode indices~--- 
the effect that may be imposed by a non-ideal fabrication process,
in particular, those of femtosecond direct laser writing~\cite{DyakonovOL}. 
In our analysis,  the perturbations are assumed to be samples of a stationary random process, characterized by an intensity and a correlation scale parameter. 
We have derived analytical equations that link these parameters with  relevant statistical characteristics  and applied them to analyse how the uncertainty evolves with field propagation length. 
We have shown that the interaction length over which the expectation value of power becomes evenly distributed among the waveguides is crucially dependent on the random process correlation scale and perturbation intensity.

\end{abstract}

\pacs{Valid PACS appear here}% PACS, the Physics and Astronomy
\maketitle

\section{Introduction}

Nowadays,  coupled waveguides have found applications  in integrated optics~\cite{Peruzzo,SzameitSUSY} and fiber signal processing schemes~\cite{Fiber}, high-power lasers~\cite{HighPowerLasers}, and as means to transform spatial field characteristics~\citep{Lens}.
With only two waveguides one can construct the  element of directional coupler (DC)~--- 
a functional analog of the traditional bulk optics beam-splitter (BS), 
and a  building block of complicated integrated optical devices.
In particular, universal multiport interferometers can be constructed using
proper arrangement of these simple elements and phase shifters~\cite{ReckUnitary,Clements},
in a way akin to constructing circuits out of lumped-parameter elements.

Although functionality of  DCs are similar to BSs,
their operating principles are different. 
Namely, while the interaction  BSs is essentially local, as it is based on scattering out of a thin layer structure,
the interaction in a DC is extended in space, since it relies on codirectional interchange of energy that occur between waves propagating in  evanescently-coupled waveguides~\cite{Snyder}.
A more general example of optical elements that derive its functionality from  the spatially extended interaction is that of 
waveguide arrays (WA), which implement coupling  between multiple waveguides simultaneously.
Owing to this unique property of coupled waveguides, they find applications 
besides just signal transforming elements,
but as an experimental research tool of optical simulations~\cite{SzameitTopology,AndersonLocalization,SilberbergSolitons}. 
In the quantum community, 
WAs have attracted intensive scrutiny,
because of  convenience in simulating
quantum systems  and performing quantum logical operations, 
provided that nonclassical states of light are used~\cite{SzameitMajorana}.
A number of suggestions on utilization of WAs have been put forward in this context, 
from  simple quantum state transform to simulation of exotic quantum systems~\cite{SzameitMajorana}
and universal quantum gate logic~\cite{Englund}.

One can construct many useful transformations with coupled waveguides having 
identical refractive index (RI) profiles.
For this, proper inter-guide distances 
are chosen to set the coupling rates parameters.
Using waveguides with mismatched RI profiles gives extra possibility to implement
a broader class of transformations and improved  functionalities~\cite{Chrostowski,Kondakci}.
However, special care should be taken at fabrication for not impose unintentional phase 
mismatches, since this degrades the targeted transformation.
This problem is especially important for waveguide structures fabricated by 
the femtosecond direct laser-writing (FSLW) technique,
which found applications for its
relative simplicity and affordability~\cite{SzameitNolte,Hirao, Fujimoto}.
The FSLW process involves several intertwined physical processes,
making the fabrication very capricious when writing two waveguides nearby.
If the distance between the waveguides are too short, then the quality of the waveguides and 
the whole functional element is degraded.
Recently, the degradation of extinction ration of  polarizing DCs created with 
FSLW in fuzed silica,  an element typically demanding very short inter-waveguide distances, 
has been attributed to uncontrolled perturbations of wavenumber mismatches 
that occur when the DC waveguides are fabricated too close to each other~\cite{DyakonovOL}.
This limit of minimal inter-waveguide distance is translated into 
the maximum possible value of coupling rates defining the miniaturization 
capabilities of the technology.

In this work, we develop a theory that describe the uncertainty of field transformation by two waveguides, 
originated 
from random perturbations of the refractive index profiles.
In this theory we assume only 
statistical knowledge about the perturbation, so that each instance of perturbations  
is a sample of a random process characterized by the perturbation power and spatial scale parameters.
The theory generalizes and broadens the one presented in~\cite{DyakonovOL}.

\section{Theory of field evolution at irregular perturbations of effective indices}

We consider field evolution in a pair of coupled waveguides.
The  core of each waveguide is represented by  RI profile
  $n_j(\vec{\rho},z)$, which is varied  with respect to propagation coordinate $z$;
here subscript $j$ marks a waveguide  ($j=1,2$).
Aside from the cross-waveguide coupling,
the field evolution is crucially influenced by the index variations. 
Assuming the waveguides to be single-mode and polarization state of the field fixed, the variations in $n_j(\vec{\rho},z)$ translates into variations of respective mode effective indices $n_j^{(eff)}(z)$~--- the parameter that will enter the equations that follows.

We split the mode effective indices into a constant part, $n_{j0}^{(eff)}$,
and  a perturbation part, $\delta{}n_{j}^{(eff)}(z)$, so that
$n_{j}^{(eff)}(z)=n_{j0}^{(eff)}+\delta{}n_{j}^{(eff)}(z)$, where
the former has the meaning of the value in the absence of perturbation,
while the latter occurs  due to an imperfect fabrication process and, therefore, can be unknown with certainty.
A plausible assumption for the waveguides fabricated by FSLW is to take perturbations small, i.e. $|\delta{}n_{j}^{(eff)}(z)|\ll{}n_{j0}^{(eff)}$,
and we set $n_{10}^{(eff)}=n_{20}^{(eff)}=n_0^{(eff)}$

Working in the context of the coupled mode theory, 
the fields in the waveguides are described by amplitudes, $a_1(z)$ and  $a_2(z)$, 
that are governed by the equations:
	\begin{equation}\label{eqn:coupled_amplitude}
	\left\{
		\begin{aligned}
			\frac{da_1}{dz}=-i\beta_1(z)a_1-iCa_2,\\
			\frac{da_2}{dz}=-iCa_1-i\beta_2(z)a_2,
		\end{aligned}
	\right.
	\end{equation}
where $C$ is the coupling coefficient and $\beta_j(z)$ is the wavenumber,
in which the effective indices enter through: 
$\beta_j(z)=kn_j^{(eff)}(z)$ ($j=1,2$),
where $k$ is the  wavenumber in the vacuum.
We note that the coupling coefficient $C$ is constant in \eqref{eqn:coupled_amplitude}, 
because of its negligible sensitivity to small variations of waveguide's parameters~---
a plausible assumption for waveguide couplers. 
\eqref{eqn:coupled_amplitude} do not take into account losses, 
so that the total optical power is conserved in the course of field evolution: $|a_1(z)|^2+|a_2(z)|^2=|a_{10}|^2+|a_{20}|^2$, 
where $a_{j0}=a_j(0)$ are the amplitude values at the input.

It is convenient to use the normalized quantities of the power difference:
$\Delta(z)=\left(|a_1(z)|^2-|a_1(z)|^2\right)/\left(|a_1(z)|^2+|a_1(z)|^2\right)$, and the
complex-valued amplitude product: $\sigma(z)=a_1^{\ast}(z)a_2(z)/\left(|a_1(z)|^2+|a_1(z)|^2\right)$, for which the equations \eqref{eqn:coupled_amplitude}
are rewritten:
	\begin{equation}\label{eqn:power_difference_system}
	\left\{
		\begin{aligned}
			\frac{d\Delta}{dz}&=2iC(\sigma-\sigma^{\ast}),\\
			\frac{d\sigma}{dz}&=iC\Delta+i\eta(z)\sigma,
		\end{aligned}
	\right.
	\end{equation}
where $\eta(z)=\beta_2(z)-\beta_1(z)=k\cdot(\delta{}n_2^{(eff)}(z)-\delta{}n_1^{(eff)}(z))$.

When  dependence $\eta(z)$ is defined as a regular function, finding solution to \eqref{eqn:power_difference_system} is not a problem.
However,
 our interest is  when information about $\delta{}n_j^{(eff)}(z)$ is uncertain, so that only  statistical distributions are known.
This incomplete knowledge may stem either from an imperfect fabrication process, 
such as in couplers created with FSLW at low inter-waveguide distance, 
or it can be implemented purposefully, 
for example,  as in the case of disordered waveguides utilized in simulations. 
As a result of this, the unknowns $\Delta(z)$ and $\sigma(z)$ possess uncertainty too, 
and therefore, \eqref{eqn:power_difference_system} is considered as a stochastic  equation set with
random parameter $\eta(z)$.

We assume that $\delta{}n_1^{(eff)}(z)$ and $\delta{}n_2^{(eff)}(z)$ are samples of 
one stationary gaussian stochastic process with the mean $\langle\delta{}n_j^{(eff)}(z)\rangle=0$  and  covariances $\langle\delta{}n_j^{(eff)}(z_1)\delta{}n_l^{(eff)}(z_2)\rangle=\alpha{}P_{jl}e^{-\alpha|z_1-z_2|}$, 
where $\alpha=1/l_0$ quantifies the scale of correlation $l_0$,
$P_{jj}$ has the meaning of the degree of uncertainty in perturbation $\delta{}n_j^{(eff)}(z)$, 
and $P_{jl}$ at $j\ne{}l$ describes the degree of cross-correlation between the perturbations in 
the waveguides; here $\langle\cdot\rangle$ stands for statistical averaging.
Therefore, we have the following model for 
stochastic process of $\eta(z)$:
	\begin{equation}\label{eqn:correlation_function}
		\begin{split}
			\langle\eta(z)\rangle&=0,\\
			\langle\eta(z_1)\eta(z_2)\rangle&=\alpha{}Pe^{-\alpha|z_2-z_1|},
		\end{split}
	\end{equation}
where  $P=P_{11}+P_{22}-2P_{12}=2(P_{11}-P_{12})$.
Due to the gaussianity of  $\eta(z)$,
the mean of the power difference, 
$\langle\Delta(z)\rangle$, and  its variance, $V_{\Delta}(z)=\langle\Delta(z)^2\rangle-\langle\Delta(z)\rangle^2$,
are enough to completely characterize the stochastic process.

There exist many ways to generate random functions $\eta(z)$ that obey 
\eqref{eqn:correlation_function}.
In particular, sampling $\eta(z)$ following to the random telegraph process is a straightforward choice.
However, this way does not suit for analytics.

It is convenient to model the stochastic process with properties \eqref{eqn:correlation_function}
as a solution to the 
differential equation: 
	\begin{equation}\label{eqn:equation_eta}
		\frac{d\eta}{dz}=-\alpha\eta(z)+\alpha\xi(z)
	\end{equation}
where $\xi(z)$ is the delta-correlated random process having
properties: $\langle\xi(z)\rangle=0$, $\langle\xi(z_1)\xi(z_2)\rangle=2P\delta(z_1-z_2)$,
with $\delta(x)$ being Dirac delta-function.
Solving~\eqref{eqn:equation_eta}, one obtains:
	\begin{equation}\label{eqn:solution_eta}
		\eta(z)=\eta_0{}e^{-\alpha{}z}+\alpha{}e^{-\alpha{}z}\int_0^{z}e^{\alpha\theta}\xi(\theta)d\theta,
	\end{equation}
where $\eta_0$ is a random quantity, that fulfills to the following:
$\langle\eta_0\rangle=0$, 
$\langle\eta_0^2\rangle=\alpha{}P$ and $\langle\eta_0\xi(z)\rangle=0$, 
in order to meet  \eqref{eqn:correlation_function}.
In particular, in the limiting case of $l_0=1/\alpha\rightarrow{}0$:
$\eta(z)=\xi(z)$.
The appearance of the delta-correlated stochastic function explicitly in 
the model of $\eta(z)$  \eqref{eqn:solution_eta}
turns out to be handy, as it enables derivation of analytical equations for stochastic characteristics.

Let us derive the equation that governs $\langle\Delta(z)\rangle$.
Firstly, notice that averaging the system \eqref{eqn:power_difference_system} 
does not change its form except for 
the
necessity to split $\langle\eta(z)\sigma(z)\rangle$ in terms of unknowns
$\langle\Delta(z)\rangle$ and $\langle\sigma(z)\rangle$.
Using \eqref{eqn:solution_eta}, it is obvious that 
$\langle\eta(z)\sigma(z)\rangle=\alpha{}e^{-\alpha{}z}\int_0^{z}e^{\alpha\theta}\langle\xi(\theta)\sigma(z)\rangle{}d\theta$,
so that now we need to calculate $\langle\xi(\theta)\sigma(z)\rangle$.
For this,  we use the property of a delta-correlated process
to have a non-zero correlation  only with the increment of a function, $\delta\sigma_{\Delta{}z}(\theta)=\sigma(\theta)-\sigma(\theta-\Delta{}z)$, accrued over 
an infinitely small interaction period $[\theta-\Delta{}z,\theta]$ prior to $\theta$:
	\begin{equation}\label{eqn:xi_sigma_corr}
		\langle\xi(\theta)\sigma(z)\rangle=\lim_{\Delta{}z\rightarrow{}0}\langle\xi(\theta)\delta\sigma_{\Delta{}z}(\theta)\rangle.
	\end{equation}
Thus, writing \eqref{eqn:xi_sigma_corr} as 
$\langle\xi(\theta)\sigma(z)\rangle=\lim_{\Delta{}z\rightarrow{}0}\int_{\theta-\Delta{}z}^{\theta}\langle\xi(\theta)\frac{d\sigma}{dz'}\rangle{}dz'$ and 
using the second equation from \eqref{eqn:power_difference_system},
the two terms have to be tackled:
$\int_{\theta-\Delta{}z}^{\theta}\langle\xi(\theta)\Delta(z')\rangle{}dz'$ and 
$\int_{\theta-\Delta{}z}^{\theta}\langle\xi(\theta)\eta(z')\sigma(z')\rangle{}dz'$.
Notice that because the integration interval is infinitely small, the terms can take non-zero values 
only if the Delta-function will be present somehow inside the integral.
Using \eqref{eqn:power_difference_system} and keeping in mind limits with respect to $\Delta{}z$ and $\Delta{}z'$, 
the first term is calculated to be:
$\int_{\theta-\Delta{}z}^{\theta}\langle\xi(\theta)\Delta(z')\rangle{}dz'=2iC\int_{\theta-\Delta{}z}^{\theta}\int_{\theta-\Delta{}z'}^{\theta}\langle\xi(\theta)\sigma(z'')\rangle{}dz'dz''+h.c.\rightarrow{}0$.
Following the same approach and using \eqref{eqn:solution_eta}, 
the term $\int_{\theta-\Delta{}z}^{\theta}\langle\xi(\theta)\eta(z')\sigma(z')\rangle{}dz'=P\langle\sigma(\theta)\rangle$,
i.e., $\langle\xi(\theta)\sigma(z)\rangle=iP\langle\sigma(\theta)\rangle$, so that the average is successfully split:
	\begin{equation}\label{eqn:eta_sigma_resolved}
		\langle\eta(z)\sigma(z)\rangle=i\alpha{}Pe^{-\alpha{}z}\int_0^{z}e^{\alpha\theta}\langle\sigma(\theta)\rangle{}d\theta.
	\end{equation}
Next, substituting \eqref{eqn:eta_sigma_resolved} into the averaged system \eqref{eqn:power_difference_system}  yields
a closed system of equations. 
In turn, the system gives the following differential equation for $\langle\Delta(z)\rangle$ alone:
	\begin{equation}\label{eqn:eq_for_delta}
		\frac{d^3\langle\Delta\rangle}{dz^3}+\alpha\frac{d^2\langle\Delta\rangle}{dz^2}+(4C^2+\alpha{}P)\frac{d\langle\Delta\rangle}{dz}+4\alpha{}C^2\langle\Delta\rangle=0,
	\end{equation}
with the initial conditions:
$\langle\Delta\rangle|_{0}=1$, 
$d\langle\Delta\rangle/dz|_{0}=0$,
$d^2\langle\Delta\rangle/dz^2|_{0}=-4C^2$.

Deriving equation for $\langle\Delta(z)^2\rangle$ is 
not so straightforward as for
 $\langle\Delta(z)\rangle$.
Using \eqref{eqn:power_difference_system},
we  write a series of relations:
	\begin{equation}\label{eqn:system_for_binaries}
		\begin{aligned}
			\frac{d\langle\Delta^2\rangle}{dz}&=4iC\langle\Delta(\sigma-\sigma^{\ast})\rangle,\\
			\frac{d\langle\Delta(\sigma-\sigma^{\ast})\rangle}{dz}&=2iC\langle\Delta^2\rangle+\\
			&+2iC\langle(\sigma-\sigma^{\ast})^2\rangle+i\langle\eta(\sigma+\sigma^{\ast})\Delta\rangle,\\
			\frac{d\langle(\sigma-\sigma^{\ast})^2\rangle}{dz}&=\frac{d\langle\Delta^2\rangle}{dz}+2i\langle\eta(\sigma^2-\sigma^{\ast{}2})\rangle,\\
			\frac{d\langle\sigma^2+\sigma^{\ast{}2}\rangle}{dz}&=\frac{1}{2}\frac{d\langle\Delta^2\rangle}{dz}+2i\langle\eta(\sigma^2-\sigma^{\ast{}2})\rangle.
		\end{aligned}
	\end{equation}

The averages $\langle\eta(\sigma+\sigma^{\ast})\Delta\rangle$ and $\langle\eta(\sigma^2-\sigma^{\ast{}2})\rangle$ have to be split to
move further in derivation.
Using \eqref{eqn:solution_eta} and \eqref{eqn:correlation_function},
and following the same approach as with derivation of \eqref{eqn:eta_sigma_resolved},
we arrive at the expression:
	\begin{equation}\label{eqn:split1}
		\langle\eta(z)\sigma(z)\Delta(z)\rangle=\frac{\alpha{}P}{8C}e^{-\alpha{}z}\int_0^{z}e^{\alpha\theta}\frac{d\langle\Delta^2\rangle}{d\theta}d\theta,
	\end{equation}
	\begin{equation}\label{eqn:split2}
		\langle\eta(z)\sigma^2(z)\rangle=2i\alpha{}Pe^{-\alpha{}z}\int_0^ze^{\alpha\theta}\langle\sigma^2(\theta)\rangle{}d\theta.
	\end{equation}
\begin{figure}
    \centering
    \includegraphics[width=0.45\textwidth]{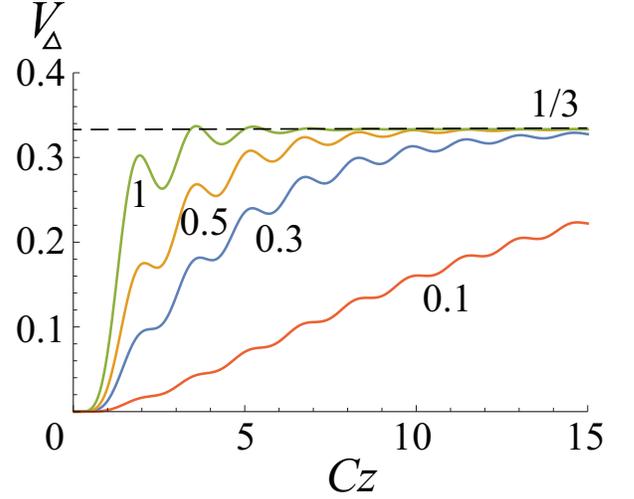}
    \caption{Covariance $V_{\Delta}$ as a function of normalized interaction length $Cz$ in the case of small-scale correlation $\alpha\rightarrow{}\infty$. The curves correspond to different values of fluctuations intensity $P$ (indicated on curves).}
    \label{fig:fig1}
\end{figure}

Finally, by differentiating the first expression in~\eqref{eqn:system_for_binaries} four times, while simultaneously using the other expressions together with \eqref{eqn:split1} and \eqref{eqn:split2},
the equation for $\langle\Delta^2\rangle$ reads:
	\begin{equation}\label{eqn:equation_quadratic_delta}
		\begin{split}
		\frac{d^5\langle\Delta^2\rangle}{dz^5}&+2\alpha\frac{d^4\langle\Delta^2\rangle}{dz^4}+(\alpha^2+4\alpha{}P+16C^2)\frac{d^3\langle\Delta^2\rangle}{dz^3}+\\
		&+\alpha(5\alpha{}P+32C^2)\frac{d^2\langle\Delta^2\rangle}{dz^2}+\\
		&+4\alpha(\alpha{}P^2+4\alpha{}C^2+12PC^2)\frac{d\langle\Delta^2\rangle}{dz}+\\
		&+48\alpha^2PC^2\left(\langle\Delta^2\rangle-\frac{1}{3}\right)=0,
		\end{split}
	\end{equation}
The initial conditions for~\eqref{eqn:equation_quadratic_delta} are:
$\langle\Delta^2\rangle|_{0}=1$, 
$d\langle\Delta^2\rangle/dz|_{0}=0$,
$d^2\langle\Delta^2\rangle/dz^2|_{0}=-8C^2$,
$d^3\langle\Delta^2\rangle/dz^3|_{0}=0$,
$d^4\langle\Delta^2\rangle/dz^4|_{0}=8C^2(16C^2+\alpha{}P)$.

It is illustrative to consider special cases. 
In the limit of small correlation scale when $\alpha\rightarrow{}\infty$,
\eqref{eqn:eq_for_delta} and \eqref{eqn:equation_quadratic_delta} are simplified:

    \begin{equation}
        \frac{d^2\langle\Delta\rangle}{dz^2}+P\frac{\langle\Delta\rangle}{dz}+4C^2\langle\Delta\rangle=0,
    \end{equation}
	\begin{equation}
	    \begin{aligned}
		    \frac{d^3\langle\Delta^2\rangle}{dz^3}+5P\frac{d^2\langle\Delta^2\rangle}{dz^2}+4(P^2+4C^2)\frac{d\langle\Delta^2\rangle}{dz}+\\
		    +48PC^2\left(\langle\Delta^2\rangle-\frac{1}{3}\right)=0.
		\end{aligned}
	\end{equation}

Fig.~1 shows the dependence of covariance $V_{\Delta}(z)$  of power difference as a function of normalized interaction length $Cz$
at different values of  the fluctuation power  in the case of small-scale correlations, $\alpha\rightarrow{}\infty$.
As can be seen from the figure, as interaction length grows, the covariance is saturated at value $1/3$, which is a witness of uniform distribution.
Indeed, for an uniformly-distributed random process $x$ defined in the range from $-1$ to $1$ the average $\langle{}x^2\rangle=\int_{-1}^{1}x^2dx=1/3$.
Also, the figure suggests that the more intensive the fluctuations is,
the faster the covariance reaches the limiting value $1/3$.

\section{Conclusion}

We have developed a theory of field transformation by a pair of coupled waveguides  that takes into account fluctuations in the mode index mismatches.
The theory can be useful in analyzing  mechanisms that degrade the functionality of integrated optics elements, especially those created with FSLW.
The approaches used in this work can also be applied to studying waveguide devices based on multiple coupled waveguides.

\section{Funding Information}

Russian Science Foundation (RSF) No 17-72-10255.


\begin{thebibliography}{1}

	\bibitem{DyakonovOL} I.V.~Dyakonov, M.Yu.~Saygin, I.V.~Kondratyev, A.A.~Kalinkin, S.S.~Straupe, and S.P.~Kulik, Laser-written polarizing directional coupler with reduced interaction length, Opt. Lett. \textbf{42}, No 20, 4231-4234 (2017).


	\bibitem{Peruzzo} R.J.~Chapman, M.~Santandrea, Z.~Huang, G.~Corrielli, A.~Crespi, M.-H.~Yung, R.~Osellame, A.~Peruzzo, Experimental perfect state transfer of an entangled photonic qubit, Nat. Commun. \textbf{7}, 11339 (2016).
	
	\bibitem{SzameitSUSY} M.~Heinrich, M.-A.~Miri, S.~St\"{u}tzer, R.~El-Ganainy, S.~Nolte, A.~Szameit, D.N.~Christodoulides, Supersymmetric mode converters, Nat. Commun. \textbf{5}, 3698 (2014).
	 
	 
	\bibitem{Fiber} B.K.~Nayar, D.R.~Smith, Optical directional coupler using polarization maintaining monomode fiber, Opt. Lett. \textbf{8}, 543 (1983).	 
	 
	 \bibitem{HighPowerLasers} J.K.~Butler, D.E.~Ackley, D.~Botez, Coupled-mode analysis of phase-locked injection-laser arrays, Appl. Phys. Lett. \textbf{44}, 293-295 (1984).
	 
	 \bibitem{Lens} P.B.~Catrysse, V.~Liu, S.~Fan, Complete power concentration into a single waveguide in large-scale waveguide array lenses, Sci. Reps., \textbf{4}, 6635 (2014).
	 
	 
	\bibitem{ReckUnitary} M.~Reck, A.~Zeilinger, Experimental realization of any discrete unitary operator, Phys. Rev. Lett. \textbf{73}, No 1, 58-61 (1994).
	
	\bibitem{Clements} W.R.~Clements, P.C.~Humphreys, B.J.~Metcalf, W.S.~Kolthammer, I.A.~Walmsley, Optimal design for universal multiport interferometers, Optica \textbf{3}, No 12, 1460-1465 (2016).
	
	
	\bibitem{Snyder} A.~Snyder and J.~Love, {\it Optical waveguide theory} (Chapman 
	\& Hall, 1983).	
	
	
	\bibitem{SzameitTopology} J.M.~Zeuner, M.C.~Rechtsman, Y.~Plotnik, Y.~Lumer, S.~Nolte, M.S.~Rudner, M.~Segev, A.~Szameit, Observation of a topological transition in the bulk of a non-hermitian system, Phys. Rev. Lett. \textbf{115}, 040402 (2015).
	
	
	\bibitem{AndersonLocalization} Y.~Lahini, A.~Avidan, F.~Pozzi, M.~Sorel, R.~Morandotti, D.N.~Christodoulides, Y.~Silberberg, Anderson localization and nonlinearity in one-dimensional disordered photonic lattices, Phys. Rev. Lett. \textbf{100}, 013906 (2008).
	
	\bibitem{SilberbergSolitons} H.S.~Eisenberg, Y.~Silberberg, Discrete spatial optical solitons in waveguide arrays, Phys. Rev. Lett. \textbf{81}, No 16, 3383-3386 (1998).	
	
	
	\bibitem{SzameitMajorana} R.~Keil, C.~Noh, A.~Rai, S.~St\"{u}tzer, S.~Nolte, D.G.~Angelakis, A.~Szameit, Optical simulation of charge conservation violation and Majorana dynamics, Optica \textbf{2}, No 5, 454-459 (2015). 	
	

	\bibitem{Englund} Y.~Lahini, G.R.~Steinbrecher, A.D.~Bookatz and D.~Englund, Quantum logic using correlated one-dimensional quantum walks, Nat. Quant. Inf. \textbf{4}, No 2, (2018).
	
	\bibitem{Chrostowski} Z.~Lu, H.~Yun, Y.~Wang, Z.~Chen, F.~Zhang, N.A.F.~Jaeger, L.~Chrostowski, Broadband silicon photonics directional coupler using asymmetric-waveguide based phase control, Opt. Express \textbf{23}, No 3, 3795-3808 (2015).
	
	\bibitem{Kondakci} H.E.~Kondakci, A.~Szameit, A.F.~Abouraddy, D.N.~Christodoulides, B.E.A.~Saleh, Sub-thermal to super-thermal light statistics from a disordered lattice via deterministic control of excitation symmetry, Optica \textbf{3}, No 5, 477-482 (2016).
	
	\bibitem{SzameitNolte} A.~Szameit, S.~Nolte, Discrete optics in femtosecond-laser-written photonic structures, J. Phys. B: At. Mol. Opt. Phys. \textbf{43}, 163001 (2010).
	
	\bibitem{Hirao} K.M.~Davis, K.~Miura, N.~Sugimoto, K.~Hirao, Writing waveguides in glass with a femtosecond laser, Opt. Lett. \textbf{21}, No 21, 1729-1731 (1996).
	
	\bibitem{Fujimoto} K.~Minoshima, A.M.~Kowalevicz, E.P.~Ippen, J.G.~Fujimoto, Fabrication of coupled mode photonic devices in glass by nonlinear femtosecond laser materials processing, Opt. Express \textbf{10}, No 15, 645-652 (2002).
	
\end{thebibliography}
\end{document}